\documentclass[11pt,a4paper]{article}

\usepackage[utf8]{inputenc} 
\usepackage{xcolor}

\usepackage{graphicx}
\usepackage{soul}
\usepackage{authblk}

\title{Ground truth? Concept-based communities versus the external classification of physics manuscripts}

\author[1,2]{Vasyl Palchykov}
\author[1]{Valerio Gemmetto}
\author[1]{Alexey Boyarsky}
\author[1]{Diego Garlaschelli}
\affil[1]{Lorentz Institute for Theoretical Physics, Leiden University, Niels Bohrweg 2, 2300RA, Leiden, The Netherlands}
\affil[2]{Institute for Condensed Matter Physics, Svientsitskii str. 1, 79011, Lviv, Ukraine}

\begin{document}
\maketitle

\begin{abstract}
Community detection techniques are widely used to infer hidden structures within interconnected systems. 
Despite demonstrating high accuracy on benchmarks, they reproduce the external classification for many real-world systems with a significant level of discrepancy.
A widely accepted reason behind such outcome is the unavoidable loss of non-topological information (such as node attributes) encountered when the original complex system is represented as a network.
In this article we emphasize that the observed discrepancies may also be caused by a different reason: the external classification itself.
For this end we use scientific publication data which i) exhibit a well defined modular structure and ii) hold an expert-made classification of research articles.
Having represented the articles and the extracted scientific concepts both as a bipartite network and as its unipartite projection, we applied modularity optimization to uncover the inner thematic structure.
The resulting clusters are shown to partly reflect the author-made classification, although some significant discrepancies are observed. 
A detailed analysis of these discrepancies shows that they carry essential information about the system, mainly related to the use of similar techniques and methods across different (sub)disciplines, that is otherwise omitted when only the external classification is considered.
\end{abstract}

\section*{Introduction}
A conflict between two members of a relatively small university organization that happened more than 40 years ago \cite{Zachary1977} has attracted a lot of attention in the scientific community so far \cite{Newman2012}.
A confrontation during the conflict resulted in a fission of the organization, known as Zachary's karate club, into two smaller groups, gathered around the president and the instructor of the club, respectively.
Predicting the sizes and compositions of the resulting factions, given the structure of the social interaction network before the split, attracted a lot of attention. This puzzle, supplemented by the known outcome, makes this system among the best studied benchmarks to test community detection algorithms \cite{Fortunato2010}.
Having verified a high level performance on the aforementioned system and on other benchmarks \cite{Lancichinetti2008}, community detection algorithms have then been massively applied to uncover tightly connected modules within large real-world systems. 
This allowed scientists to identify, for instance, Flemish- and French-speaking communities in Belgium using mobile phone communication networks \cite{Blondel2008}, detect functional regions in the human or animal brain from neural connectivity \cite{Bullmore2009}, observe the emergence of scientific disciplines \cite{Shibata2008} and investigate the evolution of science using citation patterns and article metadata \cite{Herrera2010,Rosvall2010,Chen2010}.

A bird's eye view on the identified clusters in real-world systems certifies their meaningfulness. 
However, an in-depth  quantitative validation of the community structure requires its comparison with an external classification of the nodes, which is accessible only for a limited number of large systems. 
Examples include crowd-sourced tag assignments for software packages \cite{Hric2014}, product categories for Amazon copurchasing networks \cite{Leskovec2007}, declared group membership for various online social networks \cite{Backstrom2006,Mislove2007} and publication venues for coauthorship networks in the computer science literature \cite{Backstrom2006}. 
Surprisingly, significant discrepancies have been identified between the extracted grouping of nodes and their external classification for these systems~\cite{Hric2014,Yang2015}. 
This message remains robust independently of the system under investigation and the technique used to uncover its community structure, and calls for a detailed inspection of such discrepancies in order to understand the reasons behind them.

One of the possible reasons concerns the strong simplification that occurs during the projection of the original complex system into a network.
This projection may omit some crucial information that cannot be encoded into the structural connection pattern \cite{Hric2014}.
The missing information may correspond to age or gender of individuals in social networks \cite{Palchykov2012,Kovanen2013} or geographical position of the nodes within spatially embedded systems \cite{Expert2011}.
Following this direction, several algorithms \cite{Bothorel2015,Newman2015} have been developed in order to handle specific nodes attributes, beside the usual connectivity patterns. 
Such approaches have been shown to identify groups of nodes that more closely reproduce the external classification in real-world systems \cite{Newman2015} than the techniques that rely on the connectivity patterns only.

In this article we argue that, independently of the aforementioned issue, the supposedly poor performance of community detection algorithms may be caused by the external classification itself and its misinterpretation. 
For instance, a system may possess several alternative classification schemes, such as thematic and methodological groupings in a system of scientific publications or in academic coauthorship networks \cite{Girvan2002}.
In such situation, the discrepancies between the community detection results and a single accessible classification (e.g. based on thematic similarity) may carry, instead, meaningful information (e.g. about methodological similarity), therefore providing an added value to the system understanding.

In this article we explore this idea by performing a detailed analysis of a scientific publication record system.
This system may be simplified to structural network representation, where the nodes correspond to scientific articles, and the links represent the relationship between them.
There are various possibilities to map these relationships: direct citation \cite{Waltman2012}, cocitation and bibliographic coupling \cite{Boyack2010} or content related similarities \cite{Boyack2011,Glenisson2005}.
Here we focus on the latter, considering scientific terms or concepts that appear within the articles.
Performing community detection on the corresponding network, we compare the results with an expert made classification of these articles, considering both similarities and discrepancies between the two different partitions.
Then we investigate the main reasons causing the most notable deviations.

This article is organized as follows. In the \texttt{Data} Section we present the dataset used; in \texttt{Methods} we introduce the methodology used to build the networks, extract the partitions and compare them with the external classification.
Finally, in \texttt{Results} and \texttt{Conclusions} we present our findings and discuss them.

\section*{Data}
We investigate a collection of scientific manuscripts submitted to e-print repository \texttt{arXiv} \cite{arxiv} during the years 2013 and 2014.
During the submission process, the authors are requested to categorize a manuscript according to the \texttt{arXiv} classification scheme by assigning at least one category to it.
Multiple-category assignments are likely to reflect the author(s) decision that a single category is not enough to fully characterize the scope of an article. 
Below we will employ a network partitioning approach that ascribes each article to a single community or cluster. 
Therefore we will be focussed only on the articles that have been assigned to a single category by the author(s).
Moreover, we will limit our investigation to the field of physics \cite{arXiv_categories2013} and consider the sets of articles submitted during the years 2013 and 2014 separately.
The restrictions towards single year datasets exclude the possible issues related to the temporal evolution of research disciplines.
The resulting datasets consist of $36386$ articles submitted during 2013 and $41848$ articles submitted during 2014, and will be referred below (together with the extracted article contents) as the \texttt{arxivPhys2013} and \texttt{arxivPhys2014} datasets, respectively.

The content of each article is represented as a set of scientific concepts, i.e. specific words (or combinations of them) that have been identified within the article full text by the \texttt{ScienceWISE.info} platform.
ScienceWISE is a web service connected to the main online repositories such as \texttt{arXiv}, \texttt{CERN} Document Server \cite{CERN} and \texttt{Inspire} \cite{Inspire}, whose peculiarity is a bottom-up approach in the management of scientific concepts. The initially created scientific ontology was followed by a continuous editing by the users, for instance by  adding new concepts, definitions and relationships.
This crowd-sourced procedure leads to the most comprehensive vocabulary of scientific concepts in the domain of physics.
Such vocabulary includes generic physics concepts like \texttt{mass} or \texttt{energy}, or more specific ones like \texttt{community detection}. 
The number $k$ of concepts may significantly vary among the manuscripts, reaching up to $k_{\rm max} \sim 400$ for  review articles. 
The average number of identified concepts $\langle{k}\rangle$ per article, together with some other characteristics of the datasets \texttt{arxivPhys2013} and \texttt{arxivPhys2014}, are shown in Tab.~\ref{tab_0}.
These datasets are accessible for download in Supplementary Information.
\begin{table}[h!]
\begin{center}
\begin{tabular}{lrrrrrcc}
 \hline
			&$N$		&$V$	&$V_{\rm gen}$	&$\langle{k}\rangle$	&$L_{\rm bp}$	&$L_{\rm idf}$\\\hline
\texttt{arxivPhys2013}	&$36386$&	$12200$&	$347$&	$37$	&	$1.3\times10^6$	&	$3.3\times10^8$	\\
\texttt{arxivPhys2014}	&$41848$&	$12728$&	$344$&	$38$	&	$1.6\times10^6$	&	$4.5\times10^8$	\\
 \hline
\end{tabular}
\end{center}
\caption{Basic characteristics of the datasets: total number of articles ($N$), total number of identified concepts ($V$) and the number of generic ones ($V_{\rm gen}$) among them; $\langle{k}\rangle$ gives the average number of non-generic concepts within arbitrary chosen article. The number of links in a bipartite representation of the data ($L_{\rm bp}$) is two orders of magnitude smaller than the number of links in its one mode projection ($L_{\rm idf}$). This results in significant differences in computational resources needed to perform community detection analysis.}
\label{tab_0}
\end{table}

\section*{Methods}
The considered datasets may be naturally encoded into bipartite networks, whose different types of nodes represent scientific articles and concepts, respectively. The unweighted links in the simplest case reflect the appearance of a concept within the article.
Alternatively, the dataset may be represented as a unipartite network with only one type of nodes that correspond to articles.
This network may be considered as a one-mode projection of the above bipartite network to the article space.
Here nodes $i$ and $j$ are connected by a link if the corresponding articles share at least a single common concept. The resulting networks are extremely dense, covering almost one half of all possible network connections (see Tab.\ref{tab_0}), thus making the unweighted network approach for unipartite networks to be rather insufficient. For this reason, we define link weights that reflect the level of content similarity between two articles, i.e. the overlap between the respective lists of concepts. Different concepts, however, may contribute differently to the similarity among two articles. Indeed, sharing a widely used concept should affect the similarity between two articles differently than sharing a specific one, suggesting that specific concepts should have a higher impact on the similarity. Each concept $c$ in the dataset is therefore weighted according to its occurrence, which may be 
accounted for by the so-called $idf(c)$ factor \cite{Jones1973}:
\begin{equation}
idf(c) = \log\frac{N}{N(c)}.
\end{equation}
Here $N$ is the total number of articles and $N(c)$ is the number of articles that contain concept $c$.
Among the $V$ concepts identified by ScienceWISE, we will consider only the specific ones, discarding the $V_{\rm gen}$ generic ones (like mass or energy, which have the corresponding label in the dataset and can thus be easily excluded in our analysis).

The content of each article can be therefore expressed by means of a ($V-V_{\rm gen}$)-dimensional concept vector $\vec{v}_i$. The element $v_{ic}$ of the concept vector of the article $i$ has non-zero value equal to $idf(c)$ only if the concept $c$ appears within the article $i$ and equals zero otherwise. 

The similarity between the contents of two articles $i$ and $j$, and the link weight $w_{ij}$ between the corresponding nodes, may then be estimated by the cosine similarity between the two concept vectors  $\vec{v}_i$ and $\vec{v}_j$ as follows:
\begin{equation}
 w_{ij} = \frac{\vec{v}_i\cdot\vec{v}_j}{|\vec{v}_i| |\vec{v}_j|}.
\end{equation}

The resulting one-mode projection network will be referred below as the \texttt{idf} representation of the data, while the original unweighted bipartite network as the \texttt{bp} one. 
The number of nodes ($N$) and links ($L_{\rm idf}$, $L_{\rm bp}$) of these networks are shown in Tab.~\ref{tab_0}.

In order to find a unipartite network partition, we will maximize a modularity function \cite{Newman2004}. To deal with bipartite networks, we adopt a co-clustering approach \cite{Blei2003} and Barber's generalization of modularity \cite{Barber2007}.

In both cases, we assume that each article may belong to a single cluster only, hence exploiting the notion of non-overlapping communities. Furthermore, the co-clustering approach makes stronger restrictions on a bipartite partition, compared to a unipartite one. Indeed, the resulting 
clusters of a bipartite partition consist of both articles and related concepts, and we assume that each concept belongs to a single cluster as well. Such restriction may be relaxed, for instance by using alternative ways to generalize modularity for bipartite network \cite{Guimera2007}.
However, we will consider co-clustering of bipartite networks since it allows us to straightforwardly employ the same greedy optimization algorithm \cite{Blondel2008} for the networks of both types.

The restriction towards a single algorithm is also caused by the result \cite{Hric2014} that i) the selected algorithm is among the ones that perform best on real-world networks and ii) the major influence on the accuracy is related to the dataset itself rather than the algorithm.
Due to the stochastic origin of this algorithm, it has been applied 100 times for unipartite networks and 1000 times for bipartite ones (due to significantly different number of links and, therefore, the required computational resources). Among the detected partitions, for each network we will select the single partition that corresponds to the highest value of modularity; this partition will be referred below as the optimal partition for each network.

\section*{Results}
A partition of a bipartite network consists of clusters that contain both articles and scientific terms (concepts), while clusters of a unipartite network partition consist of articles only. 
To compare both unipartite and bipartite partitions with the external article classification, we will be focussed only on the articles that fall into each cluster. 
Thus, by referring below to a cluster of bipartite partition we mean the set of articles that belong to the specified cluster.
In this perspective, the external classification of the articles is represented by the \texttt{arXiv} standard split into different subject classes or categories (\texttt{astro-ph}, \texttt{cond-mat}, etc.).

Then, given two partitions $P$ and $Q$ of the same network (for instance a detected network partition and the  \texttt{arXiv} classification), an initial comparison between them has been performed using an information-based symmetrically normalized mutual information:
\begin{equation}\label{eq_NMI}
 I_{\rm N}(P,Q) = \frac{2I(P,Q)}{H(P)+H(Q)}.
\end{equation}
Here $I(P,Q)$ is the mutual information \cite{Meila2007} between two partitions $P$ and $Q$, and $H(P)$ is the entropy of partition $P$.
The normalized mutual information $I_{\rm N}(P,Q)$ may vary between $0$ and $1$. A value of $0$ indicates that the two partitions have no information in common, while a value of $1$ corresponds to identical partitions.
In Tab.~\ref{tab_1} we show the level of similarity between each optimal partition and the \texttt{arXiv} classification ones.
\begin{table}[h!]
\begin{center}
\begin{tabular}{lcc}
 \hline
    &	\texttt{bp}&	\texttt{idf}\\\hline
\texttt{arxivPhys2013}	&	$0.58$&	$0.54$\\
\texttt{arxivPhys2014}	&	$0.60$&	$0.55$\\
 \hline
\end{tabular}
\end{center}
\caption{Similarity between network partitions and external classification: normalized mutual information $I_{\rm N}$ (\ref{eq_NMI}) between the optimal partition of each network representation and \texttt{arXiv} classification of the articles. Both \texttt{bp} and \texttt{idf} partitions demonstrate similar value of similarity to \texttt{arXiv} classification.}
\label{tab_1}
\end{table}
The reported values of normalized mutual information indicate the existence of some common information between  automatically identified clusters of articles (both in the bipartite and unipartite cases) and the author based classification.
However, the values being quite far from the possible maximum of $1$ reflect evidence for some discrepancies between the partitions.
Below we perform a detailed analysis of these discrepancies. Here we will show the results for the \texttt{arxivPhys2013} dataset; similar findings can be observed in the \texttt{arxivPhys2014} case and are shown in Supplementary Information.

\begin{figure}[!h]
\begin{center}
\includegraphics[height=7cm]{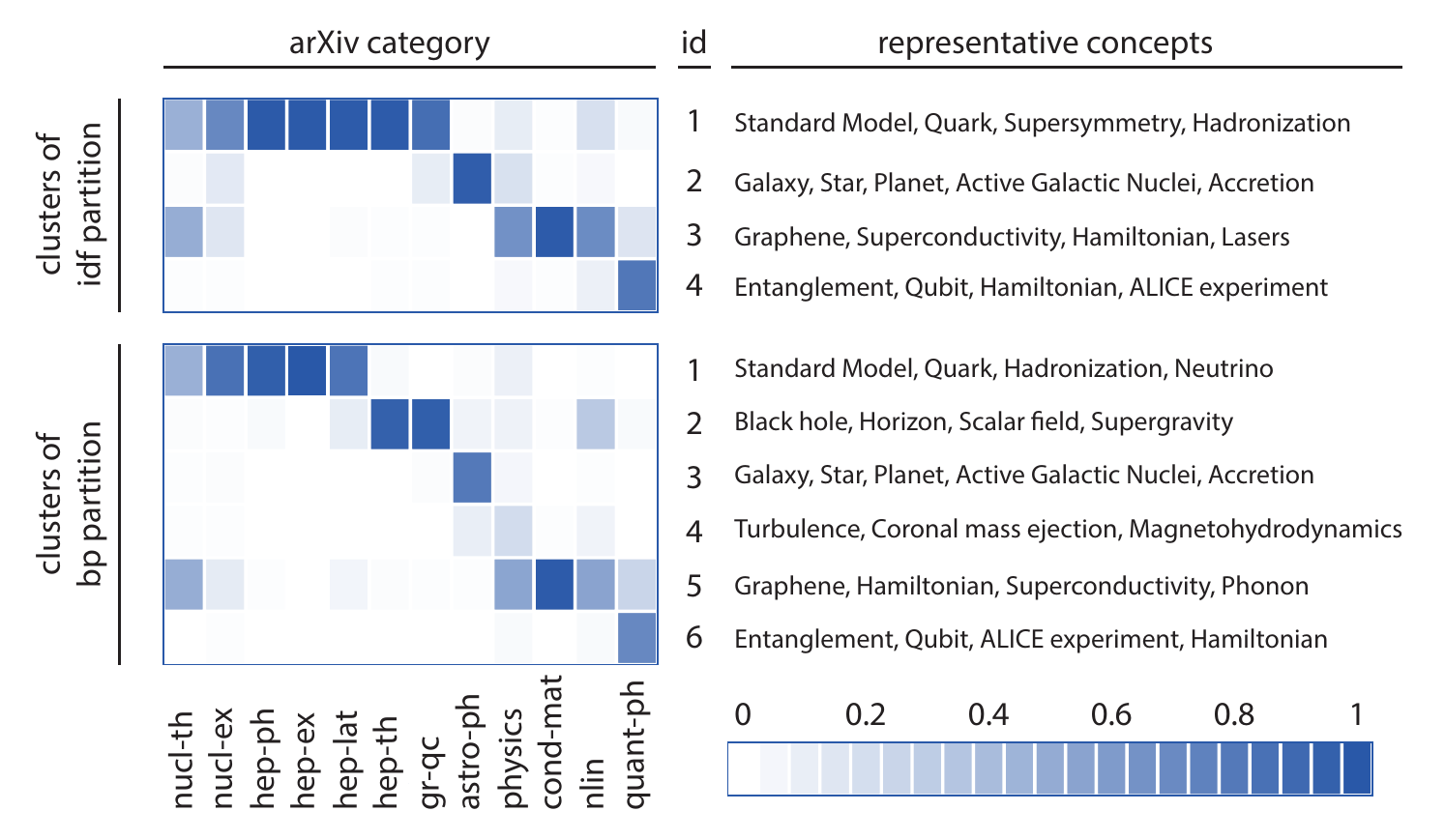}
\end{center}
\caption{ \textbf{Inner composition of arxivPhysics2013 partitions.} The color of each cell accounts for the fraction of articles of a given category belonging to a cluster (each column sums to 1). The articles of the same categories tend to incorporate into single clusters as justified by clearly visible block-diagonal structure of both \texttt{idf} and \texttt{bp} partitions. Nevertheless, the split of some categories into distinct clusters may be observed. For instance, the articles of \texttt{nucl-th} category are roughly equally split among \texttt{hep}- and \texttt{cond-mat}-dominated categories. On the right, the most representative concepts for each cluster are shown.}\label{fig_1}
\end{figure}

The first difference is observed in the numbers of detected clusters and of \texttt{arXiv} subject classes: while the number of categories in the \texttt{arXiv} classification scheme is $12$ \footnote{In fact, there are $13$ physics categories in \texttt{arXiv} classification scheme, but there is no single article in \texttt{arxivPhys2013} dataset that belong to \texttt{math-ph} category only}, the number of clusters in our partitions is only equal to $4$ in the \texttt{idf} and to $6$ in the \texttt{bp} network representations, respectively\footnote{By performing a detailed comparison we ignore all single-node clusters, which contain the articles for which no concepts has been identified.}. 
Indeed, the articles of some different \texttt{arXiv} categories tend to belong to a single cluster. This may be clearly observed in Fig.~\ref{fig_1} that shows the fraction of articles of each \texttt{arXiv} category belonging to each cluster in the resulting partitions.
This merger is especially visible for different high energy physics (\texttt{hep}) categories (\texttt{hep-ph}, \texttt{hep-ex}, \texttt{hep-lat} and \texttt{hep-th}): in the \texttt{idf} partition, almost $99\%$ of all these articles fell into a single cluster, independently of the sub-field. This result, despite deviating from the \texttt{arXiv} classification scheme, is reasonable since we observe a union of almost all papers about high energy physics, no matter if they deal with experimental or theoretical issues.

Instead, in the \texttt{bp} partition the articles of the four \texttt{hep} categories are almost entirely distributed among two clusters, focussed on experimental and theoretical issues, respectively. The first of them joins $95\%$ of all articles that belong to experimental categories (\texttt{hep-ph}, \texttt{hep-ex} or \texttt{hep-lat}), while the second one contains $94\%$ of all theoretical (\texttt{hep-th}) articles. Thus, the presence of more clusters within the bipartite network partition allows us to identify methodologically different clusters of articles within the \texttt{hep} categories, in particular dividing theoretical papers from experimental ones.

Even though the split of \texttt{hep} articles into two groups may be simply explained by the different approaches used to study the phenomena, a further result can be observed from Fig.~\ref{fig_1}: in the bipartite network partition, \texttt{hep-th} articles tend to form a single cluster with the articles that belong to general relativity and quantum cosmology (category \texttt{gr-qc}) rather than with the other high energy physics articles, thus appearing to be more similar to \texttt{gr-qc} papers rather than to the other \texttt{hep} ones. 

\begin{figure}[!h]
\begin{center}
 \includegraphics[height=7cm]{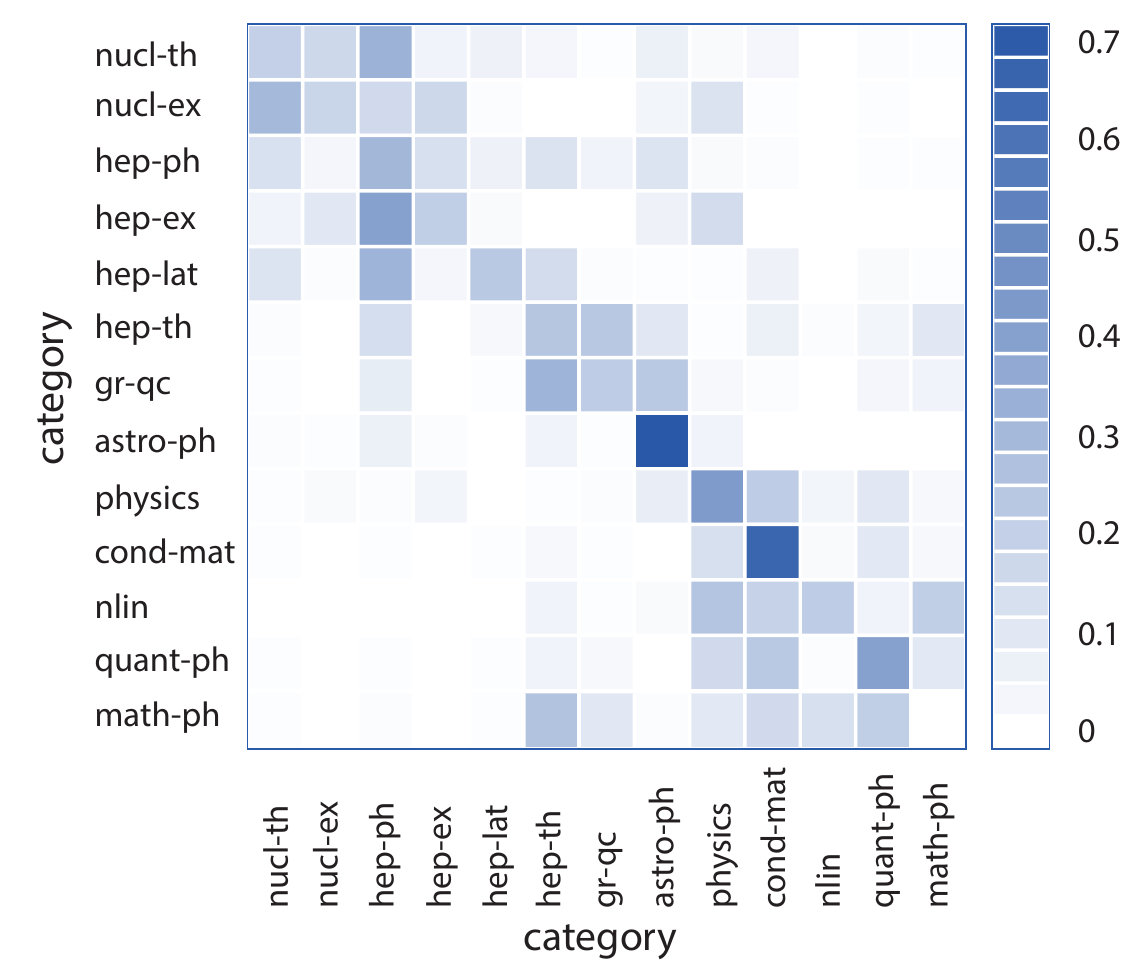}
\end{center}
\caption{\textbf{Co-occurrence matrix of \texttt{arXiv} categories during year 2013.}
Built on the complementary dataset to \texttt{arxivPhys2013}, this matrix reflects the relationships between \texttt{arXiv} categories and allows to justify the meaningfulness of some remarkable discrepancies, like the merger of \texttt{hep-th} and \texttt{gr-qc} articles.
Each non-diagonal element reflects the fraction of articles in which two specified categories have co-occurred.
The diagonal cells represent the fractions of articles that have been assigned a single category, i.e. they concerns the articles of the \texttt{arxivPhys2013} dataset.
A normalization procedure has been performed such that each row of the matrix sums to 1.}\label{fig_2}
\end{figure}

Such relatedness between the articles of the two theoretical physics categories (\texttt{hep-th} and \texttt{gr-qc}) may be verified independently by a category co-occurrence analysis. To show this, we will use the complementary part of the investigated dataset.
This set consists of all articles that have been submitted to \texttt{arXiv} during the same 2013 year, but for which the authors have assigned at least two different categories. 
Thus, no article of this set overlaps with the clustered \texttt{arxivPhys2013} collection.
Irrespective of the details of the decision-making process through which authors assign multiple categories, this multiplicity reflects the author's decision that the scope of the article can not be properly covered by a single category of a given classification scheme. Whilst several categories may cover the scope of a single research article, the co-occurrence of the same two categories in a significant fraction of articles may reflect some hidden relationships between them.
The corresponding empirical co-occurrence matrix is shown in Fig.~\ref{fig_2} and indicates the fraction of articles of a given category that have been co-submitted to the other categories.
The diagonal elements of this matrix indicate the fraction of articles of each category that have been assigned a single category by the author(s), i.e. the articles of the \texttt{arxivPhys2013} dataset.
A normalization procedure has been performed such that each column of the matrix sums to 1.

Fig.~\ref{fig_2} confirms that the \texttt{hep-th} subject class is indeed more related to the \texttt{gr-qc} class than to the other \texttt{hep} categories: \texttt{hep-th} co-occurred with \texttt{gr-qc} in $1721$ articles, and with all other \texttt{hep} categories in only $1286$ articles, even though the number of the corresponding \texttt{hep} papers (\texttt{hep-ph}, \texttt{hep-ex}, \texttt{hep-lat}) exceeds the number of \texttt{gr-qc} ones threefold.
This high level of relatedness between \texttt{hep-th} and \texttt{gr-qc} categories justifies the merging of the articles of these categories into a single cluster and indicates the meaningful deviation from the \texttt{arXiv} classification scheme. 
It is worth to mention that in the \texttt{idf} partition, where all \texttt{hep} category articles tend to belong to a single cluster, the same cluster is supplemented by $87\%$ of all \texttt{gr-qc} articles, in agreement with the result observed above. Moreover such a tendency in not restricted to the dataset for the selected year: it  has also been observed for the \texttt{arxivPhys2014} one.

The same approach explains the presence of a significant fraction of \texttt{physics}, non-linear (\texttt{nlin}) and quantum physics (\texttt{quant-ph}) articles into \texttt{cond-mat} clusters. It also allows us to understand a possible reason why nuclear physics articles (both theory and experiment) occur significantly within \texttt{hep} clusters. However, it cannot explain the presence of roughly one half of \texttt{nucl-th} articles into the condensed matter cluster (cluster No.~3 in \texttt{idf} and No.~5 in \texttt{bp} partitions) in both network representations. 
The latter deviation from the article classification, which is not explained by category co-occurrence, does not exclude that similarities between these topics exist but are considered not strong enough by the authors to label the articles with both subject classes.
To uncover the possible essence of these similarities, we examine the top representative concepts that characterize the \texttt{nucl-th} articles that belong to the two different clusters, see Table~\ref{tab_2}.
\begin{table}[h!]
\begin{center}
\begin{tabular}{|rl|rl|}
 \hline
$\%$ & Concept (cluster no.~1) & $\%$ & Concept (cluster no.~3)\\\hline
43&	Hadronization		&	55&	Isotope\\
39&	Isospin			&	53&	Hamiltonian\\
37&	Pion				&	39&	Hartree-Fock\\
33&	Degree of freedom	&	36&	Quadrupole\\
32&	Heavy ion collision	&	34&	Isospin\\
31&	Quark			&	31&	Nuclear matter\\
29&	Chirality			&	30&	Degree of freedom\\
29&	Hamiltonian		&	28&	Mean field\\
29&	Nuclear matter		&	26&	Harmonic oscillator\\
26&	Coupling constant	&	25&	Spin orbit\\
 \hline
\end{tabular}
\end{center}
\caption{Representative concepts of two groups of articles categorized as \texttt{nucl-th}. The left side of the table represents the group of articles that fell into \texttt{hep} dominated cluster (no.~1) in \texttt{idf} partition. The right side -- the other group: the \texttt{nucl-th} articles that fell into \texttt{cond-mat} dominated cluster (no.~3). For each group, the numbers next to the concepts give the percentage of articles in which the concept has been identified. The table allows us to make a suggestion that the two groups of articles significantly differ by the methods used to investigate nuclear matter.}
\label{tab_2}
\end{table}
In both cases, the top representative concepts contain the ones that characterize the object of investigation within theoretical nuclear physics, such as \texttt{Isotope}, \texttt{Isospin} or \texttt{Nuclear matter}.
However, one may clearly identify method-related concepts, such as \texttt{Hartree-Fock}, \texttt{Hamiltonian}, \texttt{Mean field} and \texttt{Random phase approximation}, among the top representative concepts of articles in the \texttt{cond-mat} cluster. These concepts clearly characterize methods that are widely used in condensed matter physics research, and that have not been identified among top concepts in any other cluster.
This result emphasizes the ability of scientific concepts found within research articles to highlight not only topics focussed on the same objects, but also methodologically similar research directions.

\section*{Conclusions}

The differences between the outcomes of community detection algorithms and possible
external classifications may have various reasons. The most notable of them
concern a possible failure of the considered algorithm or the unavoidable loss of data about real
complex systems determined by their representation as networks. To deal with the
first issue, algorithms are heavily tested on benchmarks, while the second issue is
still under investigation \cite{Newman2015}.
In this article, we emphasize a third possible reason behind such discrepancies,
i.e. the fact that the external classification itself may possess its own limitations.
For this reason we performed a detailed investigation of a scientific publication
system which i) may be naturally represented as a network and ii) owns an external
author-made classification of scientific articles. While, indeed, some discrepancies
are caused by the lack of data (for instance in the case of the articles for which
no concept has been identified), we argue that the most remarkable of them may
reflect real commonalities across different subject classes.
Academic publications are traditionally categorized and classified\footnote{Document classification and categorization are different processes: classification refers to the assignment one or more predefined categories to a document, while categorization refers to the process of dividing the set of documents into priory unknown groups whose members are in some way similar to each other \cite{Jacob2004}.} according
to objects or phenomena under investigation. The same phenomena, however, may be
explored using various approaches, experimental observation and theoretical modeling
being among them. On the other hand, the phenomena that belong to different
research topics may be investigated using the same methods, composing the core of
the interdisciplinary research. Thus, a more comprehensive classification or research
articles may be represented by a two layer categorization scheme, where one layer
reflects phenomena or objects while the other one stands for the methods of investigation.
Usually, these two layers are not taken equally into account. The expert made classification
may include rather a strong bias towards the object layer. The reasons
involve the classification scheme itself and the limited knowledge about all other research
disciplines that employ the same methods. Instead, automatic concept-based
categorization has no direct preference for any of the layers: the extracted concepts
correspond both to phenomena and methods, and the algorithm has no information
about the possible division of the concepts. Thus, the observed discrepancies may
reflect the dominance of the methodological layer over the other one, which corresponds
to phenomena or objects. Similar results have been previously observed within the
collaboration network of scientists at Santa Fe Institute \cite{Girvan2002}, where, besides the
expected grouping around common topics, some methodologically driven clusters
have been observed.

This shows that the failure in reproducing an external classification may indicate a
genuinely more complicated organization within the system, in addition to the lack
of data or algorithmic mistakes. Besides developing sophisticated algorithms to deal
with real systems, we should therefore keep in mind that some observed discrepancies
may go beyond the standard classification and carry important information
about the system under study. 
These results warn us against the use of the notion of ground truth. Indeed, it may happen that what we consider as the ground truth is just one of the possible reference points, rather than some absolute truth. Understanding the information employed to define the so-called ground truth is therefore crucial in order to perform a proper comparison between external classification and automatically retrieved communities.

\section*{Acknowledgements}
 The authors thank A.~Cardillo, A.~Martini, P. de Los Rios, O.~Ruchaiskiy and D.~Larremore for useful discussions, and A.~Magalich for preparation of the data. This work was supported by SNSF project No. 147609 Crowdsourced conceptualization of complex scientific knowledge and discovery of discoveries and by the EU project MULTIPLEX (contract 317532).

\bibliographystyle{unsrt}
\bibliography{article}

\end{document}